\documentclass[%
 aip,
 amsmath,amssymb,
 reprint,%
]{revtex4-1}

\usepackage{graphicx}
\usepackage{dcolumn}
\usepackage{bm}

\usepackage[utf8]{inputenc}
\usepackage[T1]{fontenc}
\usepackage{mathptmx}
\usepackage{upgreek}
\usepackage[utf8]{inputenc}
\usepackage[T1]{fontenc}
\usepackage{mathptmx}
\usepackage{subfigure}
\usepackage{amsmath}
\usepackage{amsfonts}
\usepackage{mathtools}
\usepackage{wasysym}
\usepackage{amssymb}
\usepackage{subfigure}
\usepackage{ragged2e}
\usepackage{upgreek}
\usepackage{graphicx}

\begin{document}

\preprint{AIP/123-QED}

\title[Thermal crosstalk of X-ray TES micro-calorimeters under FDM readout]{Thermal crosstalk of X-ray transition-edge sensor micro-calorimeters under frequency domain multiplexing readout}

\author{D. Vaccaro}
 \email{d.vaccaro@sron.nl}
\affiliation{NWO-I/SRON Netherlands Institute for Space Research, Niels Bohrweg 4, 2333CA Leiden, Netherlands}
\author{H. Akamatsu}%
\affiliation{NWO-I/SRON Netherlands Institute for Space Research, Niels Bohrweg 4, 2333CA Leiden, Netherlands}
\author{M.P.~Bruijn}
\affiliation{NWO-I/SRON Netherlands Institute for Space Research, Niels Bohrweg 4, 2333CA Leiden, Netherlands}
\author{L. Gottardi}%
\affiliation{NWO-I/SRON Netherlands Institute for Space Research, Niels Bohrweg 4, 2333CA Leiden, Netherlands}
\author{R.~den~Hartog}
\affiliation{NWO-I/SRON Netherlands Institute for Space Research, Niels Bohrweg 4, 2333CA Leiden, Netherlands}
\author{J.~van~der~Kuur}
\affiliation{NWO-I/SRON Netherlands Institute for Space Research, Landleven 12, 9747 AD Groningen, Netherlands}
\author{A.J.~van~der~Linden}
\affiliation{NWO-I/SRON Netherlands Institute for Space Research, Niels Bohrweg 4, 2333CA Leiden, Netherlands}
\author{K. Nagayoshi}%
\affiliation{NWO-I/SRON Netherlands Institute for Space Research, Niels Bohrweg 4, 2333CA Leiden, Netherlands}
\author{K.~Ravensberg}
\affiliation{NWO-I/SRON Netherlands Institute for Space Research, Niels Bohrweg 4, 2333CA Leiden, Netherlands}
\author{M.L.~Ridder}
\affiliation{NWO-I/SRON Netherlands Institute for Space Research, Niels Bohrweg 4, 2333CA Leiden, Netherlands}
\author{E.~Taralli}%
\affiliation{NWO-I/SRON Netherlands Institute for Space Research, Niels Bohrweg 4, 2333CA Leiden, Netherlands}
\author{M.~de Wit}%
\affiliation{NWO-I/SRON Netherlands Institute for Space Research, Niels Bohrweg 4, 2333CA Leiden, Netherlands}
\author{J.R.~Gao}
\affiliation{NWO-I/SRON Netherlands Institute for Space Research, Niels Bohrweg 4, 2333CA Leiden, Netherlands}
\affiliation{Optics Group, Department of Imaging Physics, Delft University of Technology, Delft, 2628 CJ, Netherlands}
\author{R.W.M. Hoogeveen}
\affiliation{NWO-I/SRON Netherlands Institute for Space Research, Niels Bohrweg 4, 2333CA Leiden, Netherlands}
\author{J.W.A.~den~Herder}
\affiliation{NWO-I/SRON Netherlands Institute for Space Research, Niels Bohrweg 4, 2333CA Leiden, Netherlands}

\date{\today}

\begin{abstract}
We have measured and characterized the thermal crosstalk in two different arrays of transition-edge sensor (TES) X-ray micro-calorimeters with frequency-domain multiplexing (FDM) readout. The TES arrays are fabricated at SRON and are a 8$\times$8 and a 32$\times$32 "kilo-pixel" uniform array. The amount of crosstalk is evaluated as the ratio between the averaged crosstalk signal and the X-ray pulse amplitudes. The crosstalk ratios (CR) for our detectors are compliant with the requirements for future X-ray space missions, such as Athena X-IFU (CR$< 10^{-3}$ for first-neighbour pixels): we measured a nearest-neighbour thermal crosstalk ratio at a level of $10^{-4}$, with a highest crosstalk ratio of $4\times 10^{-4}$ for the kilo-pixel array (worst case, center of array) and $1\times 10^{-4}$ for the 8$\times$8 array, with a margin of improvement achievable by optimizing the Cu metallization and the width of the Si supporting structures (muntins) in the backside of the TES array chip. Based on the measured crosstalk ratios, we have estimated the impact on the spectral resolution by means of noise equivalent power (NEP) considerations and a Monte Carlo simulation, finding an average degradation in quadrature of less than 40~meV, compliant with the < 0.2~eV requirement for Athena X-IFU.

\end{abstract}

\maketitle

\begin{quotation}
This paper has been accepted for publication in \textit{IEEE Transactions on Applied Superconductivity} .
\end{quotation}

\section{Introduction}

Transition-edge sensors \cite{tes} (TESs) are very sensitive devices used as bolometers or single-photon micro-calorimeters in a variety of present and future astrophysical missions, such as LiteBIRD \cite{litebird}, HUBS \cite{hubs}, Athena X-IFU \cite{athena}. Given the stringent limitations of space-borne missions in terms of available cooling power at sub-K temperatures, the readout of TESs is usually performed under a multiplexing scheme. Several multiplexing methods exist, whose means are to modulate the detector signals with a set of orthogonal carriers within a limited bandwidth and perform their transmission via a common line.

At SRON, we are developing large, close-packed arrays of TES bolometers for sub-mm and far-IR observatories and micro-calorimeters for X-ray astronomy. To perform their readout, we have been developing a frequency-domain multiplexing (FDM) technology with base-band feedback \cite{bbfb} (BBFB).

In our FDM readout for X-ray micro-calorimeters, the voltage bias for the TESs is provided by a comb of sinusoidal carriers generated by a custom warm electronics board (DEMUX). A tuned, high-Q LC bandpass filter is placed in series with each detector, limiting the information bandwidth and allowing only one frequency from the carrier comb to provide the AC bias.  An X-ray releasing energy in a TES causes its temperature and resistance to increase and its current to drop, then electro-thermal feedback (ETF) brings the system back to its original point, resulting in an amplitude-modulation of the signal. The TESs' output currents are summed at the input port of a superconducting quantum interference device (SQUID). An analogue low-noise amplifier (LNA) provides a further amplification at room temperature before the analog-to-digital conversion in the DEMUX board, where demodulation is performed to recover the X-ray signal. In the BBFB scheme, the demodulated signals are re-modulated and fed back to the SQUID via the feedback line, after compensating for the delay due to transmission time, to null the SQUID output. This effectively increases the available dynamic range and allows for a large multiplexing factor (\textit{i.e.} number of pixels per readout chain). In our case, the FDM readout is performed in the bandwidth from 1~MHz to 5~MHz with a pixel separation of 100~kHz. For more details on the scheme and the most recent demonstration of our FDM readout for X-ray TES micro-calorimeters, see the study from Akamatsu~\textit{et al.}\cite{hiroki2021}.

Given the proximity of the pixels both physically and in frequency space, the spectral resolution of the detectors is susceptible to crosstalk. We observe two types of crosstalk: thermal crosstalk and electrical crosstalk. When dealing with crosstalk between two pixels, we call the "Perpetrator" the pixel where an actual X-ray event is detected and the "Victim" the pixel where a spurious signal due to crosstalk is observed.

Thermal crosstalk (TCT) occurs because a part of the energy deposited by an X-ray in the Perpetrator pixel is not compensated via the ETF mechanism, $i.e.$ by the reduction in Joule heating dissipated by the TES. As such, the level of thermal crosstalk depends on the effective strength of ETF, quantified by the loop gain $\mathcal{L}$:

\begin{equation}
\mathcal{L} = \frac{\alpha}{n}\left[1 - \left( \frac{T_{b}}{T} \right)^{n} \right] \frac{1-R_{s}/R}{1+R_{s}/R+\beta}\ ,
\end{equation}
where $R_{s}$ and $R$ are the shunt resistance and TES resistance, $n$ is the thermal conductance exponent (for the detectors used in our experiments, $n \approx 3.2$), $T_{b}$ and $T$ are the bath temperature and TES temperature and $\alpha \equiv$~d(lnR)/d(lnT) and $\beta \equiv$~d(lnR)/d(lnI) are the TES temperature and current sensitivity, respectively. If an amount of energy $E_{0}$ is deposited by an X-ray in the Perpetrator, a fraction $E_{0}/(1+\mathcal{L})$ is not compensated by a reduction of Joule heating of the TES and dissipates into the thermal bath, becoming source of thermal crosstalk. TCT can then be minimized by proper heat-sinking the array. We usually bias our devices along the transition at $R = 20\%$ of the normal resistance, where typically $\alpha \sim 500$ and $\beta ~\sim 5$, resulting in $\mathcal{L} \sim 25$ (the value for other quantities necessary for $\mathcal{L}$ calculation are given in Section~\ref{setup}).

Electrical crosstalk (ECT) occurs when the electrical signal in the readout chain produced by the detection of an X-ray in the Perpetrator pixel causes a spurious signal to be observed in the Victim pixel. ECT does not originate from the detector array, but merely from the readout circuit. There are several factors that can produce ECT in a FDM system, such as carrier leakage, common inductance in the AC bias line and/or at the SQUID input line, or mutual inductance between neighbour inductors in the LC filter chip.

In this paper we present an experimental characterization of thermal crosstalk for two different TES arrays fabricated at SRON, using two of our FDM setups, namely the XFDM~setup and the kilo-pixel~setup (described in next section). ECT is not the main focus of the paper and is not treated.

This paper is structured as follows. In Section \ref{setup} we describe the TES arrays and the FDM setups used for the measurements. The experimental approach is explained in Section \ref{design} and the results are discussed in Section \ref{results}.

\section{Experimental setup}\label{setup}

To give a better overview on the two setups, the detector parameters and readout components described below are also summarized in Table~\ref{tablesetup}.

\subsection{TES arrays}

The detectors used in the experiments reported in this paper have been developed at SRON. The arrays used in the two setups have both different devices and different layout. The XFDM~setup hosts a $8\times 8$ uniform array, 32 TESs of which are connected to the readout circuit. The kilo-pixel~setup hosts a $32\times 32$ uniform kilo-pixel array, 38 TESs of which are connected to readout circuit.

\begin{figure}
\includegraphics[width=0.48\textwidth]{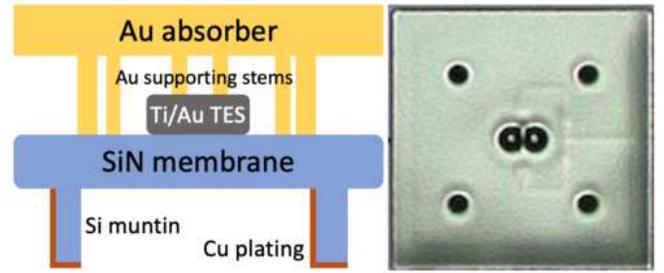}
\caption{Left: not-to-scale sketch for a single pixel. Right: microscope top-view picture of a single pixel ($240\times 240\upmu$m$^{2}$, $250\upmu$m pitch), with the gold absorber and the six supporting stems in evidence and the Nb leads for the TES readout visible underneath.}\label{tesdesign}
\end{figure}

Each TES consists of a Ti/Au bilayer deposited on a 500~nm thick SiN membrane. The bilayer dimensions are $50\times 10\ \upmu$m$^{2}$ and $80\times 20\ \upmu$m$^{2}$ for the array hosted in the XFDM and kilo-pixel~setup, respectively. The bilayer is coupled to a $240\times 240\ \upmu$m$^{2}$, 2.3~$\upmu$m thick Au absorber (heat capacity $C = 0.85$~pJ/K at 90~mK) via two central pillars, with four additional corner stems providing mechanical support, as shown in Figure~\ref{tesdesign}, right plot. The $50\times 10\ \upmu$m$^{2}$ devices have critical temperature $T_{\textup{C}} \simeq 85$~mK, normal resistance $R_{\textup{N}} \simeq 125$~m$\Omega$ and thermal conductance $G\sim 50$~pW/K at $T_{\textup{C}}$. The $80\times 20\ \upmu$m$^{2}$ devices have $T_{\textup{C}} \simeq 85$~mK, $R_{\textup{N}} \simeq 100$~m$\Omega$ and $G\sim 80$~pW/K at $T_{\textup{C}}$.

Thermalization of the TES chip is provided by Au wire-bondings between the Cu sample holder and the gold layer on top of the array. To further improve the thermal contact, a 1~$\upmu$m Cu layer with 20~nm Au capping is added to the Si supporting structures (muntins) on the backside of the chip. The width of the muntins for this batch of arrays is 60~$\upmu$m. Due to the position and orientation of the sample holder inside the evaporation chamber, the muntins are not uniformly metallized, as shown in Figure \ref{tesdesign}, left plot.

For further details about the TES array fabrication, see the study from Nagayoshi \textit{et al.} \cite{ken}. For further details about their AC operation and spectral performance, see the study from de Wit \textit{et al.} \cite{martin}.

\subsection{FDM setups}

In both the XFDM and kilo-pixel~setups, the TESs are coupled to custom LC filters \cite{marcel} and superconducting transformers for impedance matching, resulting in an effective inductance of 3.7~$\upmu$H and 2.8~$\upmu$H for the XFDM and kilo-pixel setup, respectively. The voltage bias is obtained through a 750~m$\Omega$ SMD (surface mount device) resistor and a capacitive divider with a 1:26 ratio, for an effective shunt resistance $R_{s} \simeq1$~m$\Omega$. The TES summed signals are pre-amplified at cryogenic stage via two SQUIDs (VTT model J3 as front-end and VTT model F5 as amplifier). Such "cold" components are hosted on custom oxygen-free high-conductivity (OFHC) copper and enclosed in a niobium shield. Superconducting Helmolz coils are employed to cancel out residual stray magnetic field after the cooldown. The detectors are hit with 5.9~keV X-rays produced from an $^{55}$Fe source hosted on the Nb magnetic shield. Aluminium foils are employed in front of the X-ray source to adjust the count rate to $\sim1$~count per second per pixel. A copper collimator is placed just above the TES array to avoid non-biased pixels being hit, as these lack the ETF and thus would cause thermal fluctuations in the array.

The setups are both housed in the same cryo-cooler, a Leiden Cryogenics dilution unit with a cooling power of 400~$\upmu$W at 120~mK. The setups are hanged via Kevlar wires to the mixing chamber to damp mechanical oscillations\cite{gotkevlar}. OFHC copper braids connecting the setups to the mixing chamber ensure the thermal anchoring. The setups' temperatures are controlled and monitored via a Ge thermistor anchored to the copper holder. For the measurements reported in this paper, the base temperature for the XFDM~setup is 55~mK and for the kilo-pixel~setup is 53~mK.

\begin{table}[!h]
\begin{center}
    \begin{ruledtabular}
    \begin{tabular}{ ccc }
    & \textbf{XFDM} & \textbf{kilo-pixel} \\ \hline
    TES design & 50x10 $\upmu$m$^{2}$ & 80x20 $\upmu$m$^{2}$\\
    TES array & 8$\times$8, 32 connected & 32$\times$32, 38 connected\\
    $R_{\textup{N}}$ & 125 m$\Omega$ & 100 m$\Omega$\\
    $T_{\textup{C}}$ & 85 mK & 85 mK\\
    $G$ @ $T_{\textup{C}}$ & 50 pW/K & 80 pW/K\\
    $C$ @ 90 mK & 0.85 pJ/K & 0.85 pJ/K\\
    $L_{eff}$ & 3.7 $\upmu$H & 2.8 $\upmu$H\\
    $R_s$ & 1 m$\Omega$ & 1 m$\Omega$\\
    $T_{b}$ & 55 mK & 53 mK\\
    \end{tabular}
    \end{ruledtabular}
\end{center}
\caption{Summary of the TES and readout parameters for the two FDM setups.}\label{tablesetup}
\end{table}

\section{Experimental approach}\label{design}

\begin{figure*}
\includegraphics[height=0.29\textwidth]{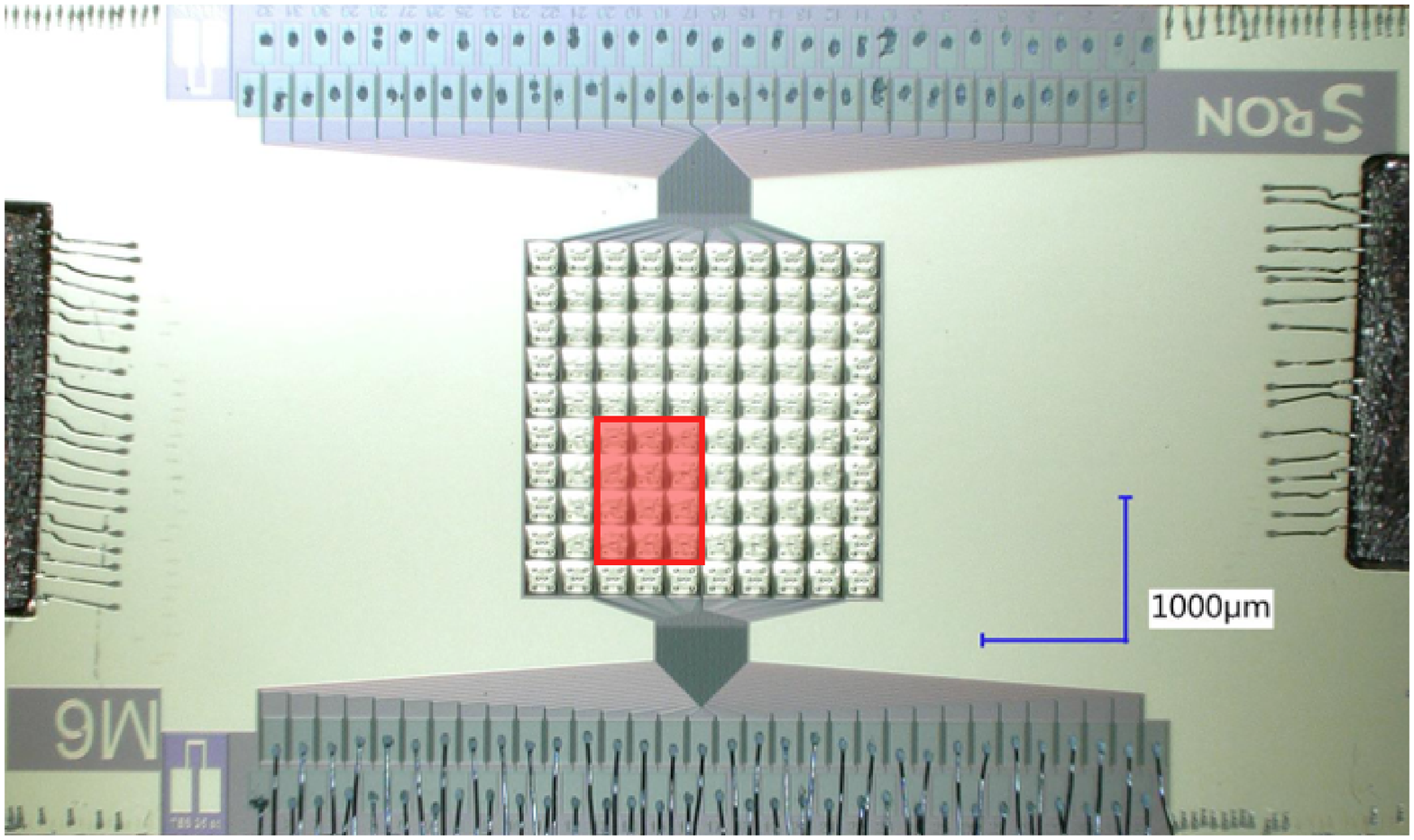}
\includegraphics[height=0.29\textwidth]{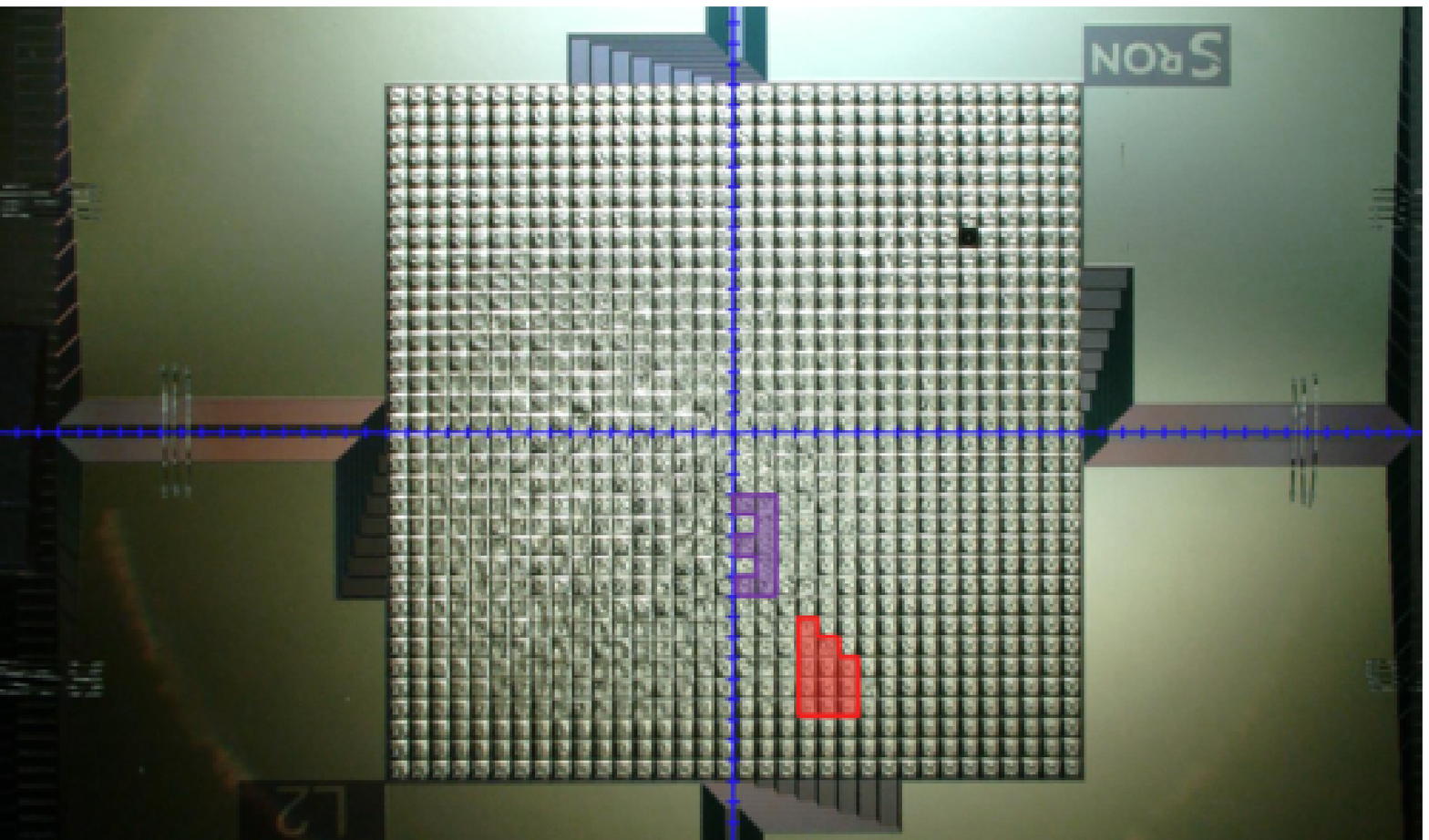}
\caption{TES arrays used in the XFDM~setup (left) and kilo-pixel~setup (right). The coloured area (purple = center of array, red = edge of array) represents the pixels activated during the crosstalk measurements.}\label{array}
\end{figure*}

To evaluate the amount of crosstalk we simultaneously activate $N$ neighbour pixels, and bias them at $\approx$~20\% of their superconducting transition, where the best spectral performances are observed. In Figure~\ref{array} we show the smaller $8\times 8$ array and the larger kilo-pixel array and the map of pixels activated during the experiments.

When one pixel is triggered by an X-ray event, data for all the active pixels is stored. The total events number is chosen to sufficiently reduce the noise floor (scaled to the amplitude of an X-ray pulse) to a level of $\approx 10^{-5}$. This number is chosen based on preliminary measurements\cite{xtalk_et} which showed the scaled amplitude of the TCT signal to be at a level of $10^{-4}$ or higher. In this way a total of $N\times 5000$ events is acquired. We can't however control the precise number of events acquired per each individual pixel.

Since we are activating multiple pixels, intermodulation distortions from non-linearities in the readout chain are expected to produce spurious line noises within the TES sensitive bandwidth. To avoid this, we use a Frequency Shift Algorithm \cite{fsapaul, fsadavide} to shift in frequency space the intermodulation lines and place them sufficiently far away from the sensitive bandwidth of the TES.

\begin{figure*}
\includegraphics[width=0.45\textwidth]{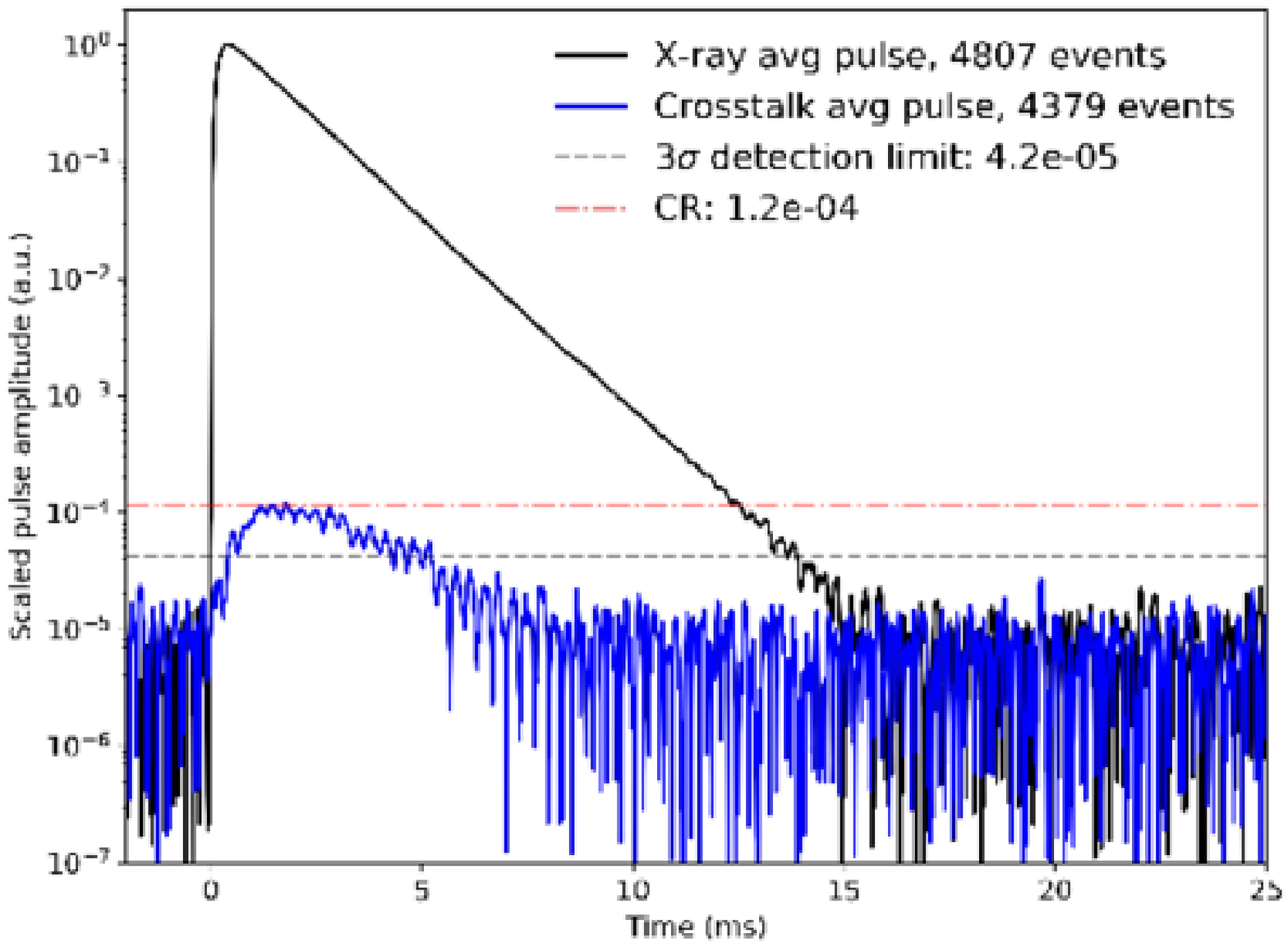}
\includegraphics[width=0.45\textwidth]{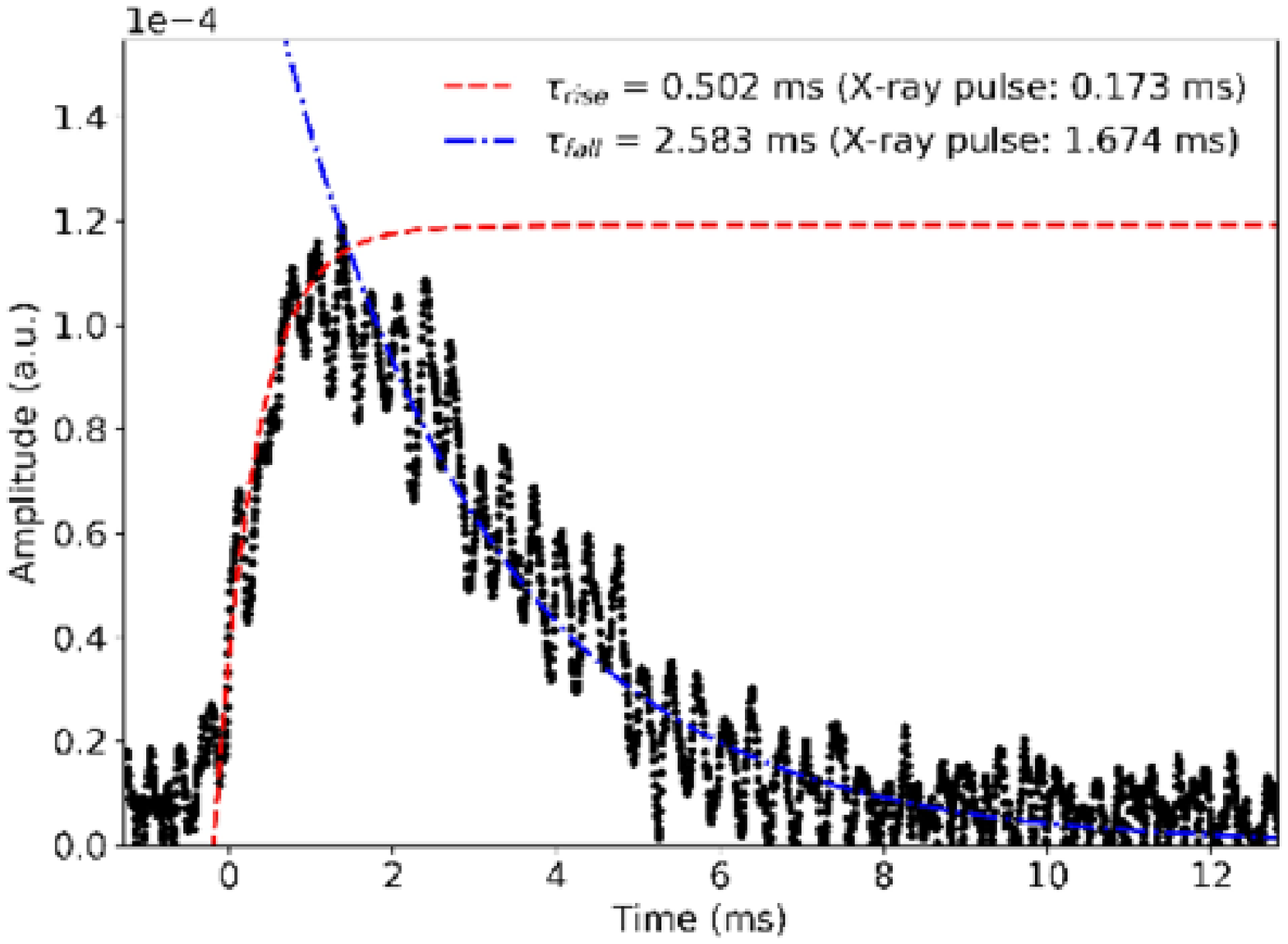}
\caption{Example of crosstalk processed data, obtained from the XFDM~setup. Left: comparison between the average 5.9 keV X-ray from the Mn-K$\alpha$ line of the $^{55}$Fe source detected by the Victim (black) and the averaged TCT signal in the Victim (blue) caused by X-rays events in a nearest neighbour Perpetrator. Right: fit of the rise time and fall time for the TCT pulse. The amplitudes are normalized to the pulse height of the average X-ray pulse. The slew rate of the X-ray pulses is at a level of 150 mA/s.}\label{crcalc}
\end{figure*}

To estimate the crosstalk ratio between the Victim pixel and the neighbouring Perpetrator pixels, we first construct the averaged X-ray pulse and TCT pulse for the Victim pixel. The average X-ray pulse is calculated by taking the average of the collected X-ray pulses, excluding those events which contain more than one pulse within the recorded event length. To analyse the data, we first subtract the baseline and take the absolute value of the amplitude to change the polarity of the pulse to positive. Then, for each Victim-Perpetrator pair, the average TCT pulse is calculated by averaging the signals for the Victim pixel acquired when the data acquisition was triggered by an X-ray in the Perpetrator pixel. The crosstalk ratio CR is then evaluated as

\begin{equation}\label{cr}
\textup{CR} = \frac{\overline{A}_{tct}}{A_{x}} ,
\end{equation}
where $\overline{A}_{\textup{tct}}$ is the amplitude of the TCT pulse averaged on 10 points ($\approx 64~\upmu$s in time scale) around the maximum and $A_{\textup{x}}$ is the peak value of the X-ray pulse. 

Both the TCT and X-ray pulse in this calculation refer to the Victim pixel. Ideally, the TCT pulse amplitude should be referred to the source X-ray pulse, which would require biasing the detectors from a perfectly uniform array on the same bias point (hence same $\alpha$ and $\beta$) and having the same gain of readout electronics. In our AC case, the bias of each TES is performed independently, which allows to tune the bias voltage individually to obtain the best performance for each pixel but, also due to the presence of weak link effect \cite{lucianowl} and the total gain of the readout electronics not being constant across the FDM bandwidth, makes it not straightforward to exactly bias to the same bias point. Therefore, calculating the crosstalk ratio using the X-ray pulse from the source pixel as reference would lead to a mis-estimation.

\begin{figure}[!h]
\includegraphics[width=0.45\textwidth]{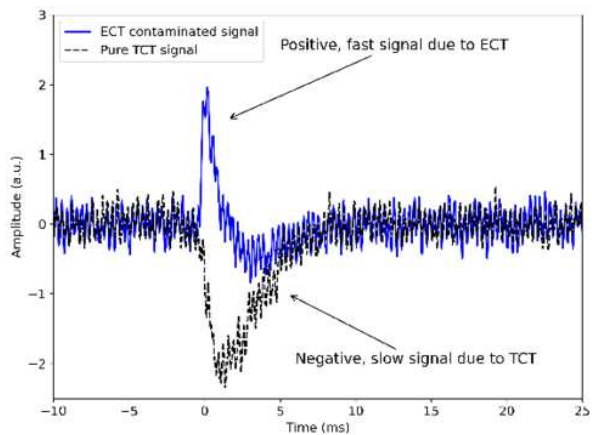}
\caption{Example of ECT contaminated signal (blue) and pure TCT signal (black dashed) for the same Victim pixel of Figure~\ref{cr_map}.}\label{ect_example}
\end{figure}

During the analysis we observed data contaminated by ECT (an example is shown in Figure \ref{ect_example}), whose signatures are (i) the opposite polarity of the signal and (ii) the faster rise time and fall time, since it is caused by the parasitic inductive coupling within the readout circuit and not by a thermo-electrical effect. In our FDM systems, the main cause of ECT is the common inductance in the AC bias and SQUID feedback line, so that the amount of ECT is proportional to $\omega/\Delta\omega$, where $\omega$ is the bias frequency of the Victim pixel and $\Delta\omega$ is the bias frequency difference between the Victim and the Perpetrator. ECT can be improved by reducing the length of the common paths in the AC bias and SQUID feedback line, consisting of lithographed strip-lines and Al bonding wires. For the results that follow, Victim-Perpetrator pairs contaminated from ECT are discarded from the analysis, so that the events analysed refer to pure thermal crosstalk.

\section{Results and discussion}\label{results}

\subsection{Crosstalk ratios}

In Figure \ref{crcalc}, left plot, we show an example of processed data for thermal crosstalk. From the noise floor level, we define a 3$\upsigma$ detection threshold, where $\upsigma$ is calculated as the root mean square of the noise floor level. We deem TCT detection and proceed with the CR calculation if two conditions are met: I) the averaged crosstalk pulse shape has the proper polarity (ECT rejection) and II) the TCT pulse amplitude is above the 3$\sigma$ detection threshold.

For Victim-Perpetrator pairs with relative distance of 2 pixels or more however, the observed TCT signals are low enough to make the 3$\upsigma$ criterion too stringent: for those pairs, condition II) is replaced with condition III) the integral of the TCT pulse is at least 30\% larger than the integral of the noise floor, where the integrals are considered in the time interval where the detected pulse signal is above the noise floor (typically few milliseconds). The threshold of 3$\sigma$ of condition II) and the use of condition III) are a compromise to effectively prevent false detection and at the same time avoiding to reject data points at pixel distances of 2 or higher, where the measured CR are small. The cost is that, for pixel distances of 2 or more, the error bar (= detection threshold $\approx 2\times10^{-5}$) on the CR value is significant.

Due to the higher ECT in the kilo-pixel~setup (due to longer wire-bonds in the AC bias and SQUID input lines), a larger number of Perpetrator-Victim pairs is excluded from the analysis than in the XFDM case. We separately fit the rising and falling flank of the TCT pulses with a single exponential function to extract the rise and fall times, as shown in Figure~\ref{crcalc}, right plot. The information about the X-ray pulse time scale is used for the Monte Carlo simulation in Section~B.2.

\begin{figure}
\includegraphics[height=0.45\textwidth]{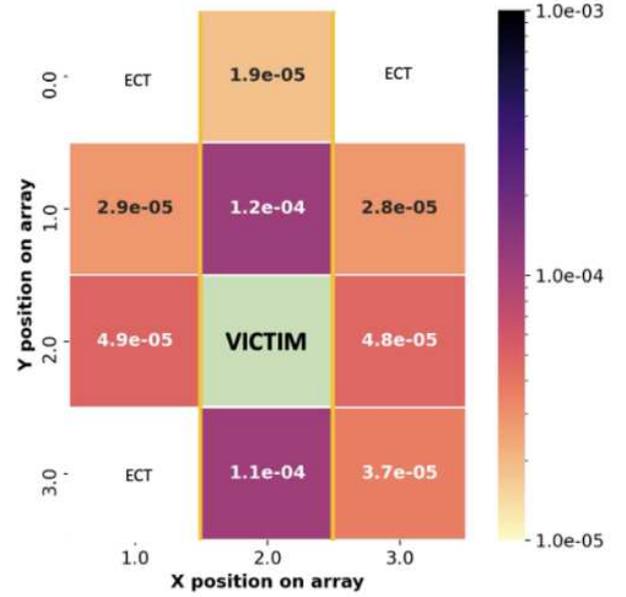}
\caption{Example of crosstalk heatmap for one Victim pixel in the XFDM~setup, constructed according to the geometrical distribution of the active pixels in the array. The numbers represent the crosstalk ratios measured for the corresponding Perpetrator pixel towards the Victim pixel. The gold-colored vertical bars in the left plot represent the area where the Cu metallization is expected to be on the silicon muntins. The color bar is scaled up to the requirement for Athena X-IFU for crosstalk between nearest neighbours ($=10^{-3}$). Perpetrators whose major source of crosstalk towards the Victim was ECT are excluded from the analysis.}\label{cr_map}
\end{figure}

\begin{figure*}
\includegraphics[width=0.49\textwidth]{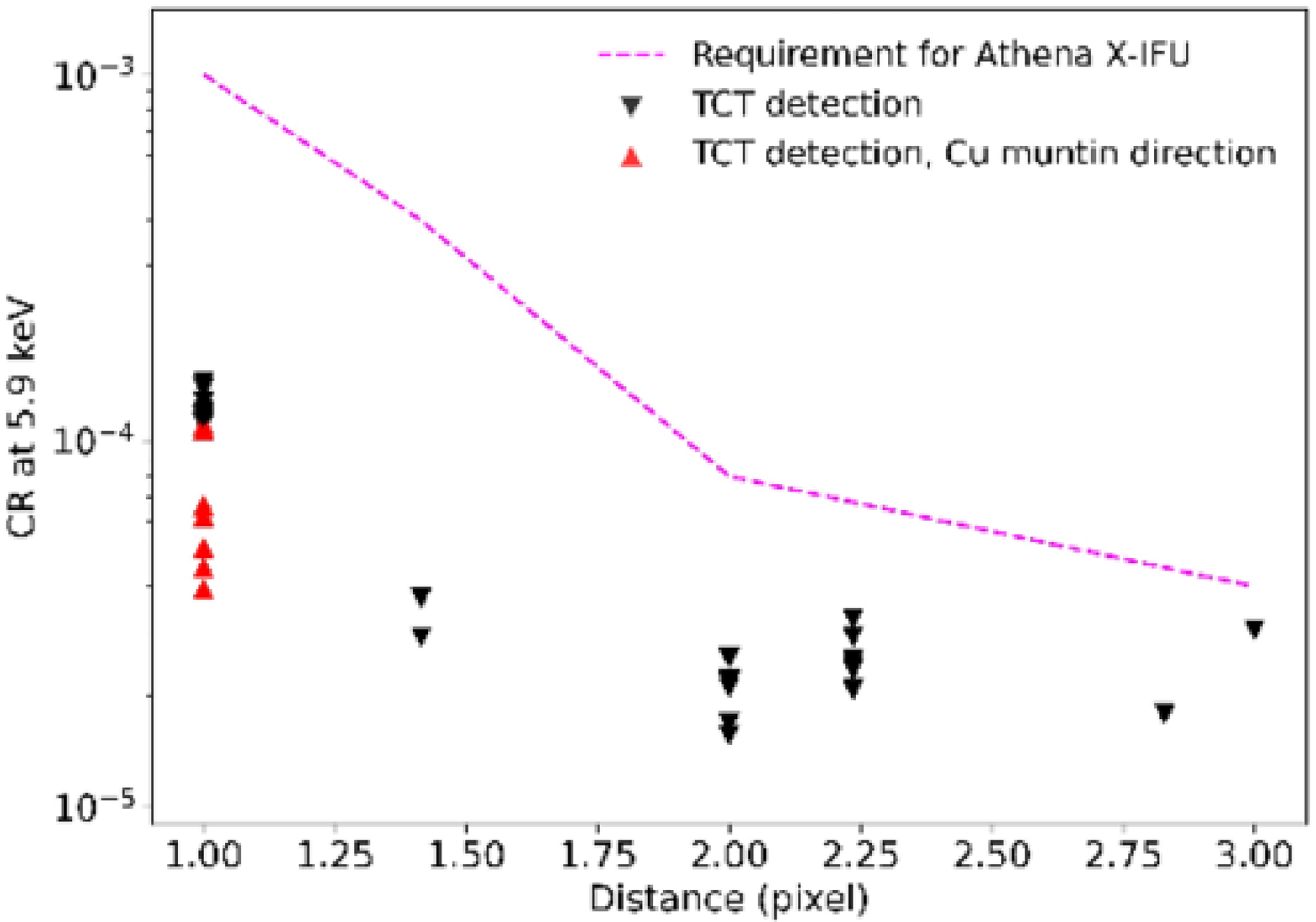}
\includegraphics[width=0.49\textwidth]{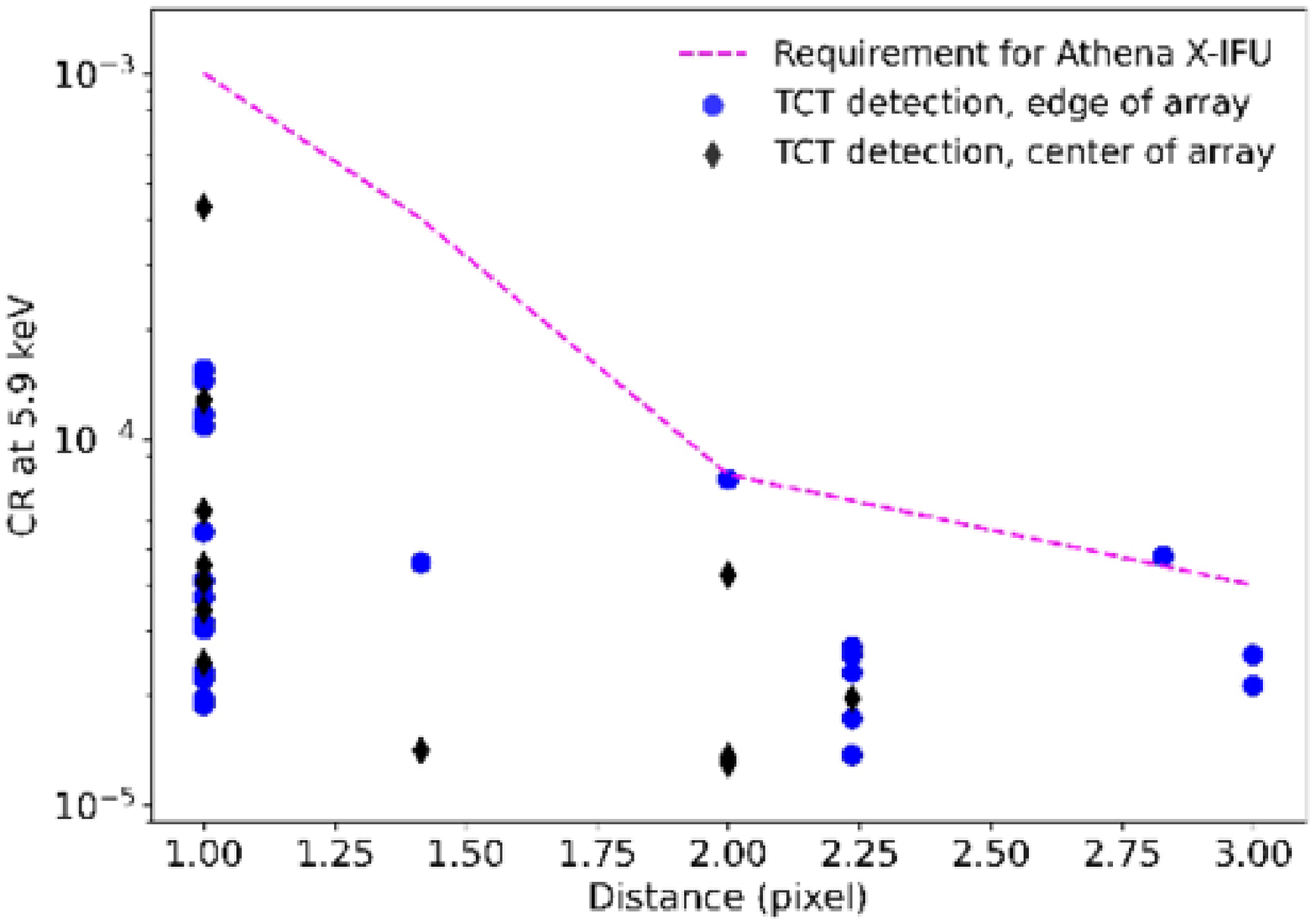}
\caption{Crosstalk ratios as a function of distance between pixels for all the Victim-Perpetrator combinations for the XFDM~setup (left) and kilo-pixel setup (right). As a reference, we compare our results with the requirement for Athena X-IFU. Data with ECT contamination is excluded. The uncertainty on CR for each data point is at a level of $2\times10^{-5}$. The error bars are not shown for plot clarity.}\label{CRdistance}
\end{figure*}

We then produce TCT maps for each pixel of the array. Figure \ref{cr_map} shows an example of such a map, as measured in the XFDM~setup (left plot) and kilo-pixel~setup (middle and right plots). We construct a map for each Victim in the array: in this way we correlate the measured CR with the relative distance between the Victim and the Perpetrator pixel, as shown in Figure~\ref{CRdistance}.

We observe a directional dependency in CR for nearest-neighbour pixels, which we interpret as the effect of the non-uniform copper metallization of the silicon muntins, which in turn depends on the position and orientation of the array on the fabrication mask. This effect is more clear for the XFDM~setup: a nearest-neighbour Perpetrator pixel lying in the Victim pixel in the direction of the Cu-plated muntin produces a TCT which is significantly lower than for Perpetrator and Victim lying in the direction on muntin without deposited copper.

Taking as reference the requirements for Athena~X-IFU\cite{miniussi}, we observe that the amount of crosstalk for our array is at acceptable levels, with highest crosstalk ratios of $1.4\times10^{-4}$ in the XFDM~setup, and for the kilo-pixel~setup $1.6\times10^{-4}$ at edge of array and $4.3\times10^{-4}$ in the center of the array. These values are expected to be improved in future batches of arrays, envisaging a larger width of the silicon muntins (from $60~\upmu$m to $85~\upmu$m) and a uniform Cu metallization.

\subsection{Impact on spectral performance}

Having measured the level of TCT for our TES arrays, we want to estimate its impact on the spectral performance of the detectors. To do so, we adopt two approaches: (1) estimation from NEP (noise equivalent power) and (2) Monte Carlo simulation.

\subsubsection{Estimation from NEP}

TCT can be considered as a disturbance in the thermal stability of the TES. In this sense,  as done in the study by Iyomoto \textit{et al.}\cite{iyomoto2}, we can estimate the impact on energy resolution via the integrated NEP, defined as

\begin{equation}
\Delta E = 2.335 \times \left( \int_{0}^{\infty} \frac{4}{\text{NEP}(f)^{2}}df \right)^{-1/2}\ ,
\end{equation}

where 2.335 is the conversion factor from RMS to FWHM, $f$ is the frequency and NEP$(f)$ is the NEP spectrum, defined as the ratio of the average noise power spectral density to the normalized detector responsivity, which we calculate from the average 6~keV X-ray pulse spectrum normalized to power units.

\begin{figure}
\includegraphics[width=0.45\textwidth]{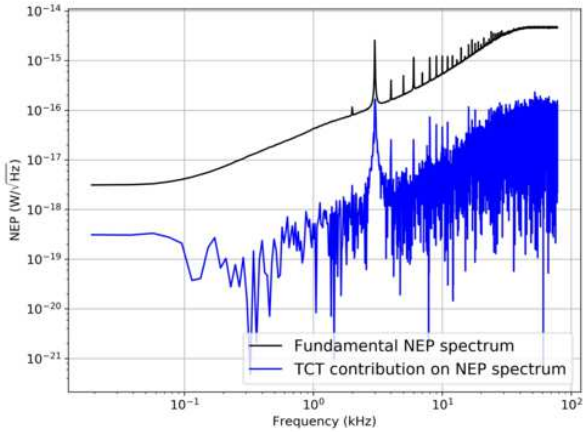}
\caption{Example of measured NEP spectrum calculated only with the fundamental noise of the detector (black) and the contribution in quadrature of the thermal crosstalk induced by a nearest neighbour pixel. The resonance at 3 kHz is a consequence of the Frequency Shift Algorithm, used to move the intermodulation distortions at 1.5~kHz and higher harmonics in the demodulated TES response. The excess noise within $G(1+\mathcal{L}_{0})/2\uppi C \simeq $~240~Hz (thermal bandwidth) in the blue curve is sign of thermal crosstalk.}\label{nepfig}
\end{figure}

We calculate the integrated NEP $\Delta E_{0}$ from the fundamental noise power spectrum, defined from the noise events acquired when no X-ray events are detected in any active TES, and then compare it to the integrated NEP $\Delta E_{\textup{TCT}}$ obtained by adding in quadrature to the fundamental noise the power spectrum of the average TCT pulse. Given the low signal-to-noise ratio of the TCT, to avoid over-estimations we first subtract the contribution of the fundamental noise from the average TCT pulse power spectrum. In Figure \ref{nepfig}, left plot, we show an example of fundamental NEP spectrum and the contribution of thermal crosstalk to the NEP spectrum. The impact on spectral performance is then estimated as

\begin{equation}\label{deg}
\sqrt{\Delta E_{\textup{TCT}}^{2} - \Delta E_{0}^{2}}\ .
\end{equation}

For nearest neighbour pixels lying along the direction with no-metallization muntins, such value is of the order of 10~meV, and is significantly lower for pixels lying along the direction of Cu-plated muntins.

This method considers the increased phonon noise in the Victim pixel caused by thermal disturbances from one Perpetrator pixel, however it is not representative for an actual FDM multi-pixel configuration, with multiple Perpetrators surrounding the Victim and an increased effective count rate. Therefore, we consider the impact on performance measured in this way as an underestimation. To make a more realistic assessment, we use a Monte Carlo simulation.

\subsubsection{Monte Carlo simulation}

\begin{figure*}
\includegraphics[width=0.48\textwidth]{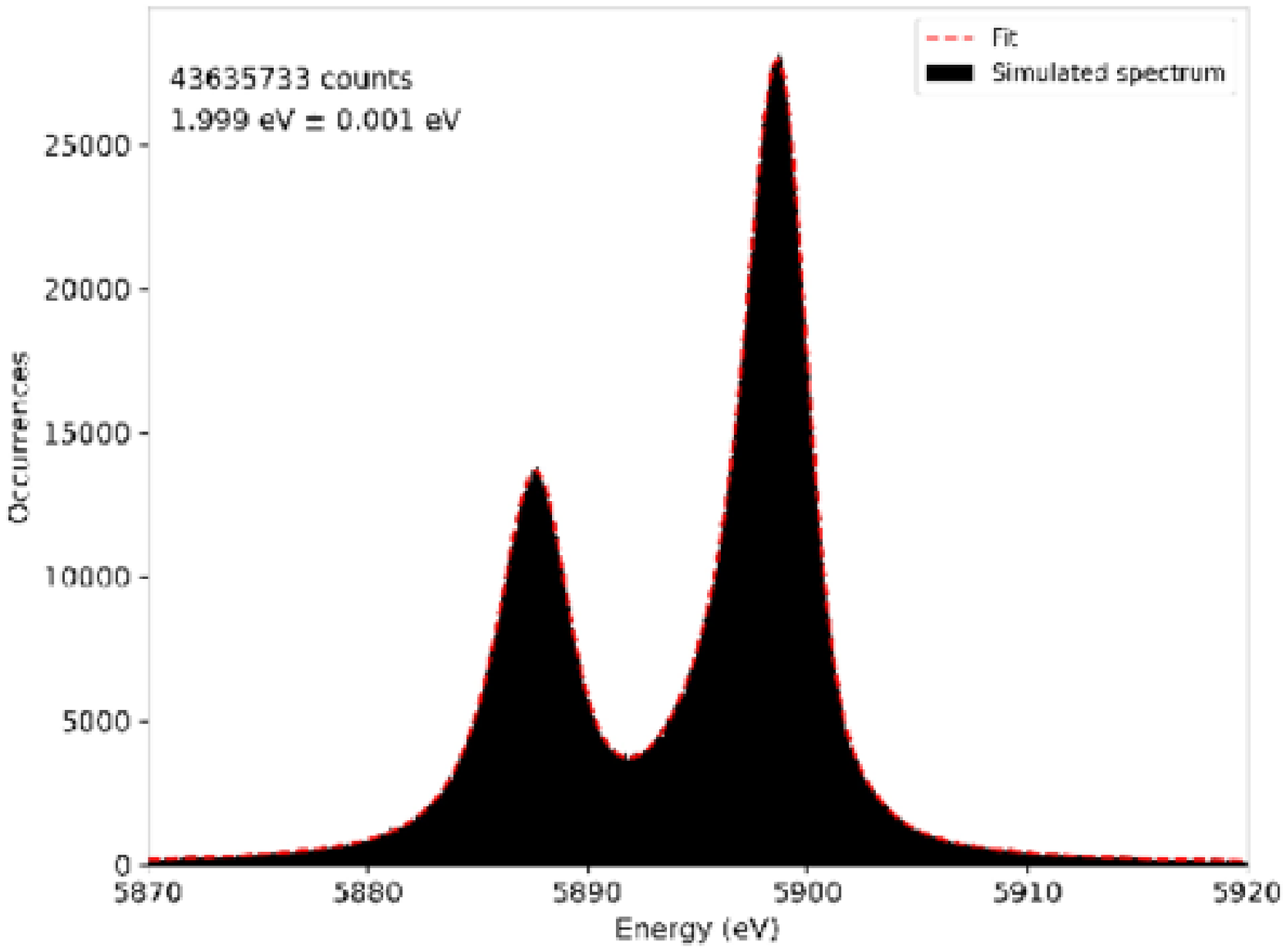}
\includegraphics[width=0.48\textwidth]{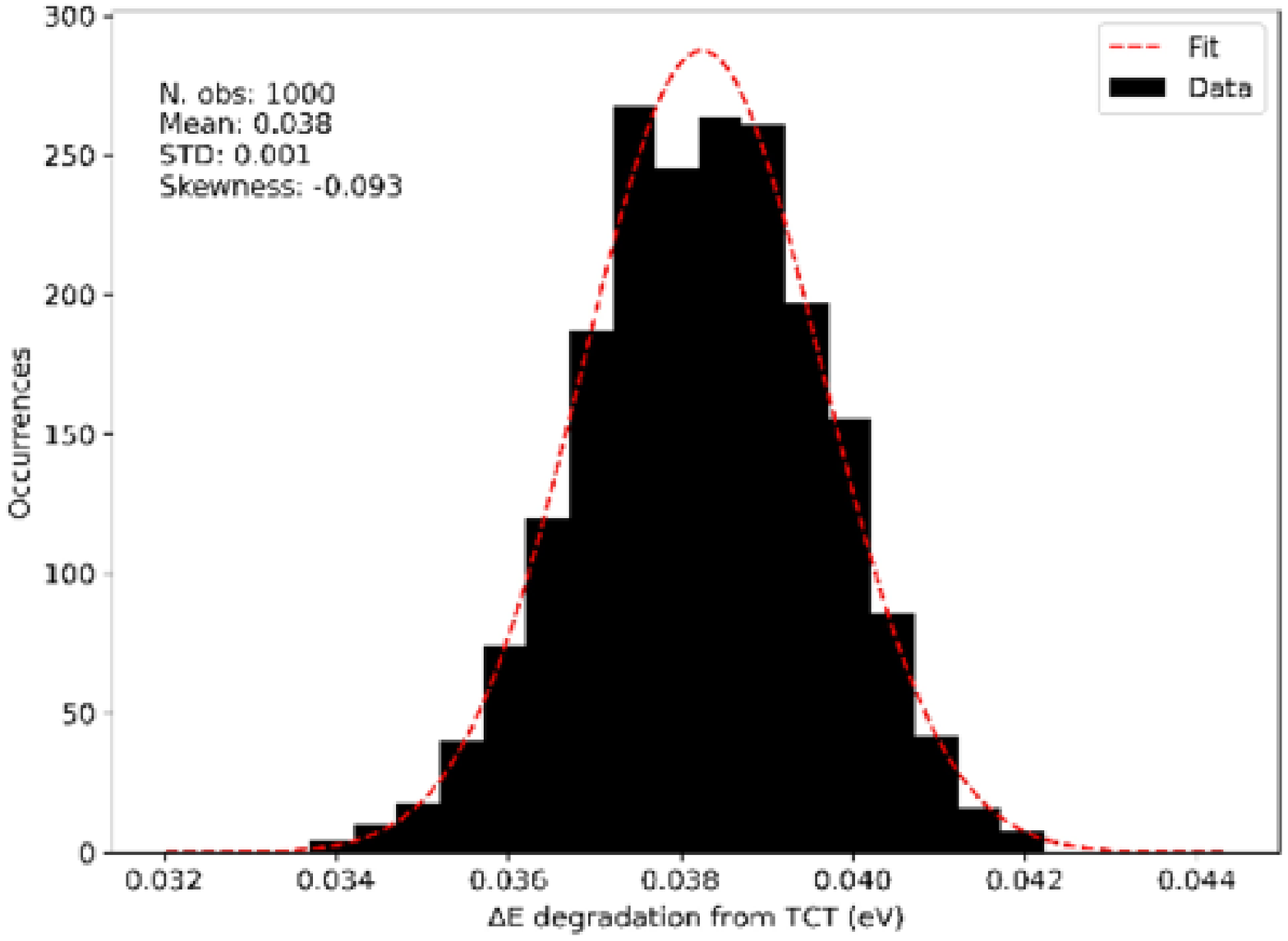}
\caption{Left: simulated energy spectrum for the Mn-K$\upalpha$ lines of a $^{55}$Fe X-ray Perpetrator. The extreme number of simulated events is necessary to reduce the statistical error to 1~meV, since from the NEP estimation we expect the impact on resolution to be of the order of at least 10~meV. Right: histogram of the degradation in quadrature caused by random TCT in the simulated spectrum for 1000 observations.}\label{MC}
\end{figure*}

In this approach, we simulate an X-ray spectrum composed by the Mn-K$\upalpha$ and Mn-K$\upbeta$ lines as typically measured from the $^{55}$Fe source in our experimental setups. To do so, we take random samples from a probability distribution constructed using the natural line data and branching ratios present in the study by Holzer \textit{et al.}\cite{holzer}, convolved with a gaussian function to tune the resolution of the K$\alpha_{1,2}$ lines to a value of 2~eV. The spectral performance is estimated by fitting the Mn-K$\alpha_{1,2}$ model to the simulated data using the Cash-statistics\cite{cstat} in the maximum-likelihood method. We simulate the energy spectrum with a number of events large enough to reduce the statistical error of the fit to 1~meV (one order of magnitude lower than the impact estimated by NEP approach), as shown in Figure~\ref{MC}, left plot.

We then contaminate the simulated spectrum with TCT events.  For the contamination to happen, an X-ray should release energy in the Perpetrator within a time window $\delta$ equal to the entire relaxation time of the Victim X-ray pulse, which we can consider to be of the order of 10~ms, as shown in Figure~\ref{crcalc}, right plot. Considering a typical count rate $\nu$ for each pixel of $\sim$~1~Hz, the probability of having an X-ray event in one pixel follows a Poisson distribution with mean $\nu\delta$. The probability of two concurrent, independent events in the Perpetrator and Victim pixel is then given by the product of the two individual probabilities. Considering that each Victim has $N_P$ neighbour Perpetrator pixels, the contamination probability is calculated as:

\begin{equation}
P_{TCT} = \left(\nu\delta e^{-\nu\delta}\right) \bigg|_{V} \sum_{N_P} \left(\nu\delta e^{-\nu\delta}\right) \bigg|_{S}  = N_P \left(\nu\delta e^{-\nu\delta}\right)^{2}\ .
\end{equation}

Considering $\delta = 10$~ms, $\nu = 1$~Hz and $N_P=8$ (assuming the TCT of higher-than-$\sqrt{2}$ neighbours to be negligible), we get $P_{TCT} \approx 7.8\times10^{-4}$. To perform a random TCT contamination, for each event of the simulated spectrum we generate a random number $X$ within 0 and 1 with a uniform probability distribution. Then, if the condition $X < P_{TCT}$ is verified, we sum the TCT contribution $\epsilon$ to the energy $E_0$ of the X-ray event. To calculate $\epsilon$ we consider the weighted average of the energy for Mn-K$\upalpha$ X-rays $\approx 5895$~eV and a CR of $10^{-4}$, so that the average energy for TCT pulse $\epsilon \approx 0.6$~eV. This is a conservative estimation, since the average CR for first neighbours is $9\times10^{-5}$ and it is expected to be even better with an improved, uniform Cu metallization.

In this way we construct a "contaminated" spectrum, which we fit to extract the energy resolution of the Mn-K$\upalpha$ lines. The degradation (in quadrature) coming from TCT is then calculated according to Eq.~\ref{deg}.

To assess the expected impact of TCT on spectral resolution, we repeat this process for a large number of observations (\textit{i.e.} random contamination of the simulated spectrum and calculation of the degradation in quadrature). In Figure~\ref{MC}, right plot, we show the results of the simulation for one thousand observations. By fitting the data with a gaussian distribution, we find an average impact of 0.038~eV~$\pm$~0.001~eV, larger than the value from NEP estimation, as expected. This value is significantly lower than the system requirements for Athena X-IFU, whose energy resolution budget for TCT is 0.2~eV, considering an instrumental energy resolution of 2.5~eV in the energy range of 0.2-7~keV for count rates per pixel of the order of 1~Hz\cite{xifu_cps}.

\section{Summary}

We presented a characterization of thermal crosstalk for two TES arrays performed in two setups for frequency-domain multiplexing readout of TES X-ray micro-calorimeters. Both the readout system and the TES arrays under study were produced at SRON. The measured level of thermal crosstalk is consistent between the two setups and is within the requirements for Athena X-IFU, with the highest crosstalk ratios for nearest neighbour pixels of $1.4\times10^{-4}$ for the 8$\times$8 array, and $4.3\times10^{-4}$ for the central region and $1.6\times10^{-4}$ at the edge of the kilo-pixel array. The crosstalk ratios appear to be well below $10^{-4}$ if the direction along the Cu muntin is considered.

By comparing the integrated NEP with and without the contribution due to the thermal crosstalk, we estimated the impact on energy resolution and the energy of the thermal crosstalk pulse for each Perpetrator-Victim combination. Using this information, we run a Monte Carlo simulation to estimate the degradation in quadrature in representative conditions for multi-pixel readout. In this way, we estimate an average impact of approximately 40~meV for a count rate per pixel of 1 Hz, compatible with the energy resolution budget allocated for thermal crosstalk in Athena X-IFU.

These results indicate that the thermal crosstalk level for our TES arrays as measured in our FDM-based system is suitable for future space-borne X-ray missions. Despite the impact of thermal crosstalk being already acceptable, we envisage the possibility to further reduce it by improving the metallization process of the silicon muntin and increasing the muntin width from 60~$\upmu$m to 85~$\upmu$m in future batches of arrays, and have a better thermalization for the TES array by using thicker or ribbon Au wire bondings.

\section*{Acknowledgements}

SRON is financially supported by the Nederlandse Organisatie voor Wetenschappelijk Onderzoek.

This work is part of the research programme Athena with project number 184.034.002, which is (partially) financed by the Dutch Research Council (NWO).

The SRON TES arrays used for the measurements reported in this paper is developed in the framework of the ESA/CTP grant ITT AO/1-7947/14/NL/BW.

\section*{Data availability}

The corresponding author makes available the data presented in this paper upon reasonable request.

\end{document}